\documentclass[12pt]{article}
 \usepackage{epsfig}
 \def\be{\begin{equation}}
 \def\ee{\end{equation}}
 \def\bea{\begin{eqnarray}}
 \def\eea{\end{eqnarray}}
 \usepackage{graphicx}

 \catcode`\@=11
 \def\lsim{\mathrel{\mathpalette\@versim<}}
 \def\gsim{\mathrel{\mathpalette\@versim>}}
 \def\@versim#1#2{\vcenter{\offinterlineskip
 \ialign{$\m@th#1\hfil##\hfil$\crcr#2\crcr\sim\crcr } }}
 \catcode`\@=12

 \catcode`@=12
\begin{document}
 \thispagestyle{empty}
 \begin{flushright}
 UCRHEP-T595\\
 Oct 2018\
 \end{flushright}
 \vspace{0.6in}
 \begin{center}
 {\LARGE \bf $U(1)_\chi$, Seesaw Dark Matter,\\ 
and Higgs Decay\\}
 \vspace{1.2in}
 {\bf Ernest Ma\\}
 \vspace{0.2in}
{\sl Physics and Astronomy Department,\\ 
University of California, Riverside, California 92521, USA\\}
\end{center}
 \vspace{1.2in}

\begin{abstract}\
It has recently been pointed out that the underlying symmetry of dark matter 
may well be $U(1)_\chi$ (coming from $SO(10) \to SU(5) \times U(1)_\chi$) with 
the dark parity of any given particle determined by $(-1)^{Q_\chi + 2j}$, where 
$Q_\chi$ is its $U(1)_\chi$ charge and $j$ its spin angular momentum. 
Armed with this new insight, previous simple models of dark matter are 
reinterpreted, and a novel idea is proposed that light seesaw dark matter 
exists in analogy to light neutrinos and is produced by the rare decay of 
the standard-model Higgs boson.
\end{abstract}

 \newpage
 \baselineskip 24pt

\noindent \underline{\it Introduction}~:~
In the decomposition of $SO(10) \to SU(5) \times U(1)_\chi$, the fermions of 
the standard model (SM) are organized into 
\begin{equation}
\underline{16} = (5^*,3) + (10,-1) + (1,-5),
\end{equation}
where
\begin{eqnarray}
&& (5^*,3) = d^c~[3^*,1,1/3,3] + (\nu,e)~[1,2,-1/2,3], ~~~ (1,-5) = 
\nu^c~[1,1,0,-5], \\ 
&& (10,-1) = u^c~[3^*,1,-2/3,-1] + (u,d)~[3,2,1/6,-1] + e^c~[1,1,1,-1], 
\end{eqnarray}
under $SU(3)_C \times SU(2)_L \times U(1)_Y \times U(1)_\chi$.  This 
symmetry breaking pattern is accomplished using the scalar $\underline{45}$ 
of $SO(10)$.  Since
\begin{equation}
\underline{45} = (24,0) + (10,4) + (10^*,-4) + (1,0)
\end{equation}
under $SU(5) \times U(1)_\chi$, a nonzero vacuum expectation value (VEV) 
of the (1,0) component will work.  Under 
$SU(3)_C \times SU(2)_L \times U(1)_Y \times U(1)_\chi$,
\begin{equation}
(24,0) = (8,1,0,0) + (1,3,0,0) + (3,2,1/6,0) + (3^*,2,-1/6,0) + (1,1,0,0).
\end{equation}
Hence a nonzero VEV of the (1,1,0,0) component will break 
$SU(5) \times U(1)_\chi$ to the SM gauge symmetry without breaking 
$U(1)_\chi$.  The Higgs scalars of the $\underline{45}$ are all superheavy 
and will not affect the phenomenology of the SM Higgs boson to be discussed 
below.   To allow the quarks and leptons to acquire mass, the scalar 
\underline{10} representation, which contains the necessary Higgs 
doublets, i.e.
\begin{equation}
\underline{10} = (5^*,-2) + (5,2),
\end{equation}
with
\begin{equation}
\Phi_1 = (\phi_1^0,\phi_1^-)~[1,2,-1/2,-2], ~~~ 
\Phi_2 = (\phi_2^+,\phi_2^0)~[1,2,1/2,2]
\end{equation}
from $(5^*,-2),(5,2)$ respectively, is required,  
resulting in the allowed Yukawa couplings
\begin{eqnarray}
d^c(u \phi_1^- - d \phi_1^0), ~~~ u^c(u \phi_2^0 - d \phi_2^+), ~~~ 
e^c(\nu \phi_1^- - e \phi_1^0), ~~~ \nu^c(\nu \phi_2^0 - e \phi_2^+),
\end{eqnarray}
as desired.  The nonzero vacuum expectation values 
$\langle \phi^0_{1,2} \rangle = v_{1,2}$ also break electroweak 
$SU(2)_L \times U(1)_Y$ to electrodynamic $U(1)$.  Since the SM gauge 
bosons all have $Q_\chi = 0$, it is obvious (but not recognized for its 
importance until recently~\cite{m18}) that all SM fermions have odd $Q_\chi$ 
and all SM bosons have even $Q_\chi$.  This means that each SM particle 
has even $(-1)^{Q_\chi + 2j}$ where $j$ is its spin angular momentum.  It is 
thus a short step to realizing that any scalar with odd $Q_\chi$ and any 
fermion with even $Q_\chi$ would have odd $(-1)^{Q_\chi + 2j}$, making it a 
natural stabilizing symmetry for dark matter.  Indeed, all previous simple 
models of dark matter based on an $Z_2$ discrete symmetry may be incorporated 
into such a framework.

The scalar \underline{126} representation of $SO(10)$ contains a singlet 
$\zeta \sim (1,-10)$ under $SU(5) \times U(1)_\chi$, which may be used   
to break $U(1)_\chi$ at the TeV scale and would allow $\nu^c$ (the 
right-handed neutrino) to obtain a large 
Majorana mass, thereby triggering the canonical seesaw mechanism for 
small Majorana neutrino masses.  This is usually described as 
lepton number $L$ breaking to lepton parity $(-1)^L$~\cite{m15}, but 
here it is clear that it has to do with the breaking of gauge $U(1)_\chi$ 
to $(-1)^{Q_\chi}$.

In the minimal supersymmetric standard model (MSSM), $R=(-1)^{3B+L+2j}$ is 
used to distinguish the SM particles from their superpartners, which 
belong thus to the dark sector if $R$ is assumed conserved.  Since $R$ 
is identical to $(-1)^{3(B-L)+2j}$, it has long been recognized~\cite{m92} 
that a theory with gauge $B-L$, broken by two units, would be a natural 
framework for dark matter.  For an incomplete list of papers on this 
topic and discussions on their relevance, see Ref.~\cite{m18}.  
In particular, the decomposition
\begin{equation}
SO(10) \to SU(3)_C \times SU(2)_L \times SU(2)_R \times U(1)_{(B-L)/2}
\end{equation}
shows that the fermionic \underline{16} of $SO(10)$ contains
\begin{eqnarray}
&& (u,d) \sim (3,2,1,1/6), ~~~ (d^c,u^c) \sim (3^*,1,2,-1/6), \\ 
&& (\nu,e) \sim (1,2,1,-1/2), ~~~ (e^c,\nu^c) \sim (1,1,2,1/2).
\end{eqnarray}
Hence $3(B-L)$ is odd for all quarks and leptons. As for the scalar sector, 
the \underline{10} representation contains the bidoublet 
$\Phi \sim (1,2,2,0)$.  Hence its $3(B-L)$ is even.  In other words, 
$(-1)^{3(B-L)}$ coincides with $(-1)^{Q_\chi}$.  However, the former requires 
a left-right intermediate scale, whereas the latter does not.  They are 
thus conceptually and phenomenologically distinct.  In this study, $U(1)_\chi$ 
separates from $SU(3)_C \times SU(2)_L \times U(1)_Y$ at the unification 
scale~\cite{m18}, 
and its symmetry breaking scale is independent of the electroweak scale.

It should also be pointed out that in  
$E_6 \to SO(10) \times U(1)_\psi$, the decomposition 
$\underline{27} = (16,-1) + (10,2) + (1,-4)$ shows that 
$Q_\psi$ may be invoked as the underlying dark symmetry as well.

\noindent \underline{\it Reappraisal of $Z_2$ Dark Matter}~:~
It has been remarked that it is very easy to invent a model of dark matter. 
The first step is to introduce a new $Z_2$ symmetry under which all SM 
particles are even and a new neutral particle of your choice is odd.  
It should then have the appropriate mass and interaction to account for 
the relic abundance of dark matter in the Universe, but not excluded 
by direct or indirect search experiments.  

The simplest such model~\cite{sz85} 
assumes a real scalar singlet, odd under $Z_2$.  It has been studied 
extensively~\cite{gambit17} and is still a viable explanation of dark matter. 
In the framework of $U(1)_\chi$, a scalar with odd $(-1)^{Q_\chi + 2j}$ 
requires it to have odd $Q_\chi$.  The scalar singlet $s \sim (1,-5)$ of 
$\underline{16}$ is such a particle.  It is in fact the scalar analog 
of $\nu^c$.  They have the same $Q_\chi$, but differ in spin.  Hence one 
is dark matter and the other is not.  In Ref.~\cite{m15}, they are both 
assigned odd lepton parity, which is now replaced by odd $\chi$ parity.
If $U(1)_\chi$ is indeed the origin of $s$, then it should be complex 
and it should have $Z_\chi$ interactions.  However, from the allowed 
$\zeta^* ss$ trilinear scalar coupling, $s$ splits into two real scalars 
with a large mass gap.  The lighter is dark matter and the heavier decays 
into the lighter plus a physical or virtual $Z_\chi$ gauge boson.
This would not affect the lighter scalar's suitability as dark matter, 
but would predict possible verifiable signatures involving $Z_\chi$. 
Note that the elastic scattering of a real scalar singlet off nuclei through 
$Z_\chi$ is forbidden because both real and imaginary parts of a complex 
scalar are needed to construct a vector current. 
The present experimental bound on $M_{Z_\chi}$ is about 4.1 TeV from 
LHC (Large Hadron Collider) data~\cite{atlas-chi-17,cms-chi-18}, which 
may be improved~\cite{kkm18} with further study.

Instead of choosing $s \sim (1,-5)$ from the \underline{16} of $SO(10)$, 
the scalar doublet $(\eta^0,\eta^-) \sim (1,2,-1/2,3)$ may also be 
considered~\cite{kkr10,kkr09,ky16}.  In that case, it is distinguished from 
$(\phi_1^0,\phi_1^-) \sim (1,2,-1/2,-2)$ and 
$(\phi_2^+,\phi_2^0) \sim (1,2,1/2,2)$ by their $Q_\chi$ charge. 
Hence $\Phi_{1,2}$ are even but $\eta$ is odd under $(-1)^{Q_\chi + 2j}$. 
This $Z_2$ discrete symmetry~\cite{dm78} allows $\eta^0$ to be dark 
matter~\cite{bhr06}, at least in principle.  However, its interaction 
with quarks through the $Z$ boson rules it out by direct-search 
experiments.  In the SM, the allowed quartic coupling 
$(\Phi^\dagger \eta)^2$ serves to split $Re(\eta^0)$ from $Im(\eta^0)$, 
and since $Z$ only couples one to the other, the offending interaction 
with quarks is avoided kinematically in elastic nuclear recoil with a 
mass gap larger than a few hundred keV.  This is known as the inert 
Higgs doublet model.  In the case of $Q_\chi$, such a 
quartic coupling is forbidden, so if $\eta$ originates from $U(1)_\chi$, 
other particles are needed for it to be dark matter.  They turn out to be 
exactly $s \sim (1,-5)$ and $\zeta \sim (1,-10)$, already discussed.  
The allowed couplings $(\eta^0 \phi_2^0 - \eta^- \phi_2^+)s$, 
$(\bar{\phi}_1^0 \eta^0 + \phi_1^+ \eta^-)s$ combined with $\zeta^*ss$ 
form the necessary effective quartic coupling as shown in Fig.~1.
In this scenario, a linear combination of $\eta^0$ and $s$ is 
dark matter.
\begin{figure}[htb]
\vspace*{-5cm}
\hspace*{-3cm}
\includegraphics[scale=1.0]{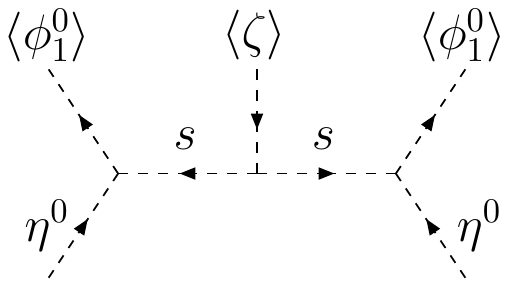}
\vspace*{-22.0cm}
\caption{Effective quartic $(\Phi_1^\dagger \eta)^2$ coupling.}
\end{figure}

Another possible simple model of dark matter is to have a singlet fermion 
$N_L \sim (1,1,0,0)$ from the $SU(5) \times U(1)_\chi$ $(24,0)$ or $(1,0)$ of 
the $\underline{45}$ of $SO(10)$.  Since $N$ has even $Q_\chi$, it is odd 
under $(-1)^{Q_\chi + 2j}$.  However, it has no renormalizable interaction with 
the particles of the SM and thus not a good dark-matter candidate without 
some additional fundamental particle such as a singlet 
scalar~\cite{lks08,bkp12} which has $Q_\chi = 0$, i.e. the scalar counterpart 
of $N$.  A more interesting option is to combine $N$ with the scalar doublet 
$\eta$ discussed in the previous paragraph because there is now an allowed 
Yukawa coupling between the left-handed lepton doublet $(\nu,l)_L$ with $N$ 
through $\eta$, i.e. $(\bar{\eta}^0 \nu_L + \eta^+ l_L)N_L$.  This forms 
the basis of the scotogenic model~\cite{m06} of radiative neutrino mass. 
Whereas the original model assumes the $(\Phi^\dagger \eta)^2$ quartic 
scalar coupling, it must now be replaced by the effective operator of 
Fig.~1.  The resulting diagram~\cite{m06-1} for generating a radiative 
Majorana neutrino mass is then given by Fig.~2.
\begin{figure}[htb]
\vspace*{-5cm}
\hspace*{-3cm}
\includegraphics[scale=1.0]{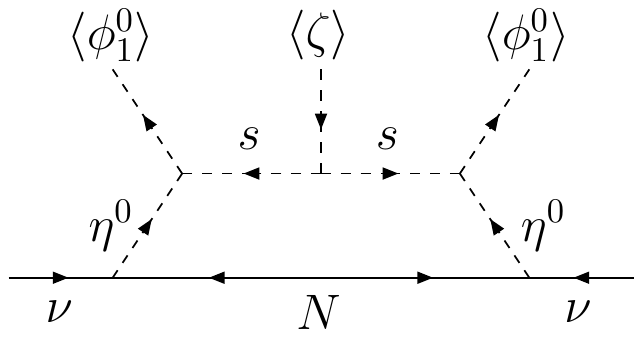}
\vspace*{-21.5cm}
\caption{Scotogenic Neutrino Mass.}
\end{figure}

Whereas $N$ could be dark matter, its only interaction with the particles 
of the SM is through the left-handed lepton doublet, and is known~\cite{kms06} 
to be restricted phenomenologically, thus limiting its viability as thermal 
dark matter.  Hence a linear combination of $s$ and $\eta^0$ is again 
the likely dark-matter candidate in this case.  They both couple to $Z_\chi$ 
but differently.  Further study is then needed to reappraise this $U(1)_\chi$ 
interpretation of the scotogenic model.

Once both $s$ and $N$ are present, the coupling $s^* N \nu^c$ is allowed. 
This has also recently been considered~\cite{b18,ck18} with the assumption that 
it is very small so that a freeze-in mechanism applies to the decay of 
$\nu^c$ to $s$ and $N$.

\noindent \underline{\it Seesaw Dark Matter}~:~
In the $U(1)_\chi$ model, the singlet neutrino $\nu^c \sim (1,-5)$ gets a 
large Majorana mass from the scalar $\zeta \sim (1,-10)$, both of which 
have even $(-1)^{Q_\chi + 2j}$.  This realizes the scenario of seesaw neutrino 
mass at the scale $\langle \zeta \rangle = u$ which may be TeV  
or higher.  Suppose the fermion singlets
\begin{equation}
N \sim (1,0), ~~~ D_\chi \sim (1,-10),
\end{equation}
from the \underline{45}, \underline{126} representations of $SO(10)$ 
are added, then the allowed Yukawa coupling $f_D \zeta^* D_\chi N$ combined 
with a large Majorana mass for $N$ would induce a small seesaw mass 
for $D_\chi$.  Note that both $N$ and $D_\chi$ have odd $(-1)^{Q_\chi + 2j}$.  
Hence $D_\chi$ could be naturally light dark matter, i.e. 
$m_{D_\chi} = f_D^2 u^2/m_N$, in parallel with the seesaw neutrino mass, i.e. 
$m_\nu = f_\nu^2 v^2/f_{\nu^c} u$. 

As for gauge $U(1)_\chi$ anomaly cancellation, the fermion 
$D^c_\chi \sim (1,10)$ from the $\underline{126}^*$ of $SO(10)$ 
should be added.   It may combine with $D_\chi$ to form a Dirac fermion, 
as proposed recently~\cite{fm18}.  Here $D^c_\chi$ is assumed to have  
an extra symmetry shared by the counterpart singlet $N^c \sim (1,0)$.  
This separate system is also assumed to be heavy and annihilate efficiently 
to SM particles through $Z_\chi$ in the early Universe. Another possible 
but different connection between seesaw neutrino mass and dark matter 
has also been proposed~\cite{betal18}, based on an imposed $Z_4$ discrete 
symmetry and a nonrenormalizable dimension-five coupling.

Consider now the interaction of $D_\chi$.  It interacts mainly with 
$Z_\chi$.  This is in analogy with $\nu$ which interacts mainly with 
$Z$ and $W^\pm$.  Just as $\nu$ decouples at a temperature of order 
1 MeV, $D_\chi$ would decouple at a temperature of order 
$T \sim 1~{\rm MeV}(m_{Z_\chi}/m_Z)^{4/3}$.  There remains however a suppressed 
Yukawa coupling to $\zeta_R = \sqrt{2}(Re(\zeta)-u)$, i.e. 
\begin{equation}
{f_D \over \sqrt{2}}{f_D u \over m_N} \zeta_R D_\chi D_\chi = 
{m_{D_\chi} \over \sqrt{2}u} \zeta_R D_\chi D_\chi.
\end{equation}
Since $m_{\zeta_R}$ is heavy, the above interaction is only realized through 
$H D_\chi D_\chi$, coming from the mixing of the SM Higgs boson $H$ with 
$\zeta$, which is itself also suppressed, i.e. of order $v/u$.  With 
these two suppressions, the resulting interaction strength will be 
very small, as shown below.

\noindent \underline{\it Higgs Decay to Dark Matter}~:~
The particles beyond the SM are the $Z_\chi$ gauge boson, the complex 
scalar $\zeta$ which breaks $U(1)_\chi$ and couples to $\nu^c \nu^c$, 
together with the $N$ and $D_\chi$ fermion singlets of Eq.~(12) which 
belong to the dark sector.  Whereas there are two Higgs doublets, i.e. 
$\Phi_{1,2}$ of Eq.~(7), one linear combination with the vacuum expectation 
value $v = \sqrt{v_1^2+v_2^2}$ is the SM analog and corresponds to the 
observed 125 GeV boson at the LHC; the other is heavier and is not 
relevant to the discussion below.

The scalar interactions between the SM Higgs $H$ and $\zeta$ is given by
\begin{equation}
V = \mu_1^2 \Phi^\dagger \Phi + \mu_2^2 \zeta^* \zeta + {1 \over 2} \lambda_1 
(\Phi^\dagger \Phi)^2 + {1 \over 2} \lambda_2 (\zeta^* \zeta)^2 + \lambda_3 
(\Phi^\dagger \Phi) (\zeta^* \zeta), 
\end{equation}
where $\Phi = (0,v+H/\sqrt{2})$ and $\zeta = u+\zeta_R/\sqrt{2}$.  The 
mass-squared matrix spanning $(H,\zeta_R)$ is then
\begin{equation}
{\cal M}^2_{H \zeta} = \pmatrix{ 2 \lambda_1 v^2 & 2 \lambda_3 vu \cr 2 \lambda_3 
vu & 2 \lambda_2 u^2}.
\end{equation}
The $H-\zeta_R$ mixing is then given by $\lambda_3 v/\lambda_2 u$.  Hence 
the $H D_\chi D_\chi$ coupling is 
\begin{equation}
f_H = {m_{D_\chi} \over \sqrt{2} u} {\lambda_3 v \over \lambda_2 u} = 
{\sqrt{2} \lambda_3 v m_{D_\chi} \over m^2_{\zeta_R}}.
\end{equation}
The decay rate $\Gamma_H$ of $H \to D_\chi D_\chi$ is then
\begin{equation}
\Gamma_H = {f_H^2 m_H \over 8 \pi} \sqrt{1-4x^2}(1-2x^2),
\end{equation}
where $x=m_{D_\chi}/m_H$.  If the reheating temperature of the Universe 
after inflation is below the decoupling temperature of $D_\chi$ for thermal 
equilibrium and above $m_H$, its only production mechanism is freeze-in 
through $H$ decay before the latter decouples from the thermal bath. 
The correct relic abundance is possible if $f_H$ is 
very small.  Hence $D_\chi$ could be FIMP (Feebly Interacting Massive 
Particle) dark matter~\cite{hjmw10}, and for $x<<1$, the right number 
density is obtained for~\cite{ac13}
\begin{equation}
f_H \sim 10^{-12} x^{-1/2}.
\end{equation}
As a numerical example which satisfies all the above conditions, let 
$m_{D_\chi}=5$ GeV, then $x=0.04$.  Assuming $\lambda_3 = 0.4$, then  
$f_H = 5 \times 10^{-12}$ in Eq.~(16) is obtained with 
$m_{\zeta_R} = 10^7$ GeV.  Assuming that this is also the value of 
$m_{Z_\chi}$, then the decoupling temperature of $D_\chi$ is about 5.2 TeV.

Since the $U(1)_\chi$ breaking scale is about $10^7$ GeV in this example of 
seesaw dark matter, the $Z_\chi$ gauge boson is much too heavy to be 
discovered at the LHC.  It is also not relevant in the thermal 
interactions of the SM particles with the dark sector below 5.2 TeV.  
Similarly, the elastic scattering of $D_\chi$ with nuclei 
through $Z_\chi$ exchange is very much suppressed, so that it is 
not detectable in direct-search experiments.

\noindent \underline{\it Concluding Remarks}~:~
Using $Q_\chi$ as a marker in $SO(10) \to SU(5) \times U(1)_\chi$ so that 
$(-1)^{Q_\chi+2j}$ distinguishes dark matter from matter, previous simple 
models of dark matter are reappraised. Furthermore, the notion is put 
forward that naturally light seesaw dark matter exists in parallel with 
naturally light seesaw neutrinos.  In the latter, the left-handed doublet 
neutrino $\nu$ couples to a heavy singlet right-handed neutrino $\nu^c$ 
through the SM Higgs doublet $\Phi$, and $\nu^c$ acquires a large Majorana 
mass through the singlet scalar $\zeta$ which also breaks $U(1)_\chi$ and 
makes $Z_\chi$ massive. As a result, $\nu$ gets a small seesaw mass. 
In the former, the fermion singlet $N \sim (1,0)$ under 
$SU(5) \times U(1)_\chi$ has an allowed large Majorana mass, whereas 
the singlet $D_\chi \sim (1,-10)$ couples to $N$ through $\zeta$, thereby 
generating a small Majorana mass for $D_\chi$.  As an example, 
$m_{\nu^c} \sim 10^7$ GeV, $m_\nu \sim 0.1$ eV, $m_N \sim 10^{14}$ GeV, 
$m_{D_\chi} \sim $ GeV may be obtained.  Note that the anchor scale 
$\langle \zeta \rangle = u$ for seesaw neutrino mass is the intermediate 
scale for seesaw dark matter.

Below the temperature of order $T \sim 1~{\rm MeV}(m_{Z_\chi}/m_Z)^{4/3}$, 
$D_\chi$ is out of thermal equilibrium with the SM particles.  However, 
there is a suppressed Yukawa interaction $f_H H D_\chi D_\chi$ which allows 
it to be produced through Higgs decay before the Universe cools below $m_H$.
It may thus be freeze-in FIMP dark matter and escape present experimental 
detection, directly or indirectly.

As for the grand unification of $SU(5) \times U(1)_\chi$, it is well-known 
that the SM particle content is inadequate for the gauge couplings to 
converge at a common mass scale.  This is however easily solved by the 
addition of new particles at intermediate scales as explicitly shown in 
Ref.~\cite{m18}.  It is also shown that it is possible to have the 
unification scale greater than $10^{16}$ GeV, thus avoiding the constraint 
from proton decay.  For each model variation considered in this paper, a full 
discussion of unification would require a similar set of new particles. 
However, the main purpose of this paper is to point out the rich physics 
possibilities of the $U(1)_\chi$ extension regarding dark matter.  Other 
possible new particles depend on the specific (but mostly arbitrary) 
scenario chosen for unification.  The 
details of how any previous proposed simple model may be fully developed 
in the $SU(5) \times U(1)_\chi$ context are left for future investigations.

\noindent \underline{\it Acknowledgement}~:~
This work was supported in part by the U.~S.~Department of Energy Grant 
No. DE-SC0008541.

\baselineskip 18pt
\bibliographystyle{unsrt}

\begin{thebibliography}{99}
\bibitem{m18} E. Ma, Phys. Rev. {\bf D98}, 091701(R) (2018) 
[arXiv:1809.03974].
\bibitem{m15} E. Ma, Phys. Rev. Lett. {\bf 115}, 011801 (2015) 
[arXiv:1502.02200].
\bibitem{m92} S. P. Martin, Phys. Rev. {\bf D46}, 2769 (1992) 
[hep-ph/9207218].
\bibitem{sz85} V. Silveira and A. Zee, Phys. Lett. {\bf 161B}, 136 (1985).
\bibitem{gambit17} For a recent review, see P. Athron {\it et al.}, 
GAMBIT Collaboration, Eur. Phys. J. {\bf C77}, 568 (2017) 
[arXiv:1705.07931].
\bibitem{atlas-chi-17} ATLAS Collaboration, M. Aaboud {\it et al.}, JHEP 
{\bf 1710}, 182 (2017) [arXiv:1707.02424].
\bibitem{cms-chi-18} CMS Collaboration, A. M. Sirunyan, A. Tumasyan 
{\it et al.}, JHEP {\bf 1806}, 120 (2018) [arXiv:1803.06292].
\bibitem{kkm18} S. J. D. King, S. F. King, and S. Moretti, Phys. Rev. 
{\bf D97}, 115027 (2018) [arXiv:1712.01279].
\bibitem{kkr10} M. Kadastik, K. Kannike, and M. Raidal, Phys. Rev. {\bf D81}, 
015002 (2010) [arXiv:0903.2475].
\bibitem{kkr09} M. Kadastik, K. Kannike, and M. Raidal, Phys. Rev. {\bf D80}, 
085020 (2009); Erratum: Phys. Rev. {\bf D81}, 029903 (2010) 
[arXiv:0907.1894].
\bibitem{ky16} T. W. Kephart and T.-C. Yuan, Nucl. Phys. {\bf B906}, 549 
(2016) [arXiv:1508.00673].
\bibitem{dm78} N. G. Deshpande and E. Ma, Phys. Rev. {\bf D18}, 2574 (1978).
\bibitem{bhr06} R. Barbieri, L. J. Hall, and V. S. Rychkov, Phys. Rev. 
{\bf D74}, 015007 (2006) [hep-ph/0603188].
\bibitem{lks08} K. Y. Lee, Y. G. Kim, and S. Shin, JHEP {\bf 0805}, 100 (2008) 
[arXiv:0803.2932].
\bibitem{bkp12} S. Baek, P. Ko, and W.-I. Park, JHEP {\bf 1202}, 047 (2012) 
[arXiv:1112.1847].
\bibitem{m06} E. Ma, Phys. Rev. {\bf D73}, 077301 (2006) [hep-ph/0601225].
\bibitem{m06-1} E. Ma, Annales Fond. Broglie {\bf 31}, 285 (2006) 
[hep-ph/0607142].
\bibitem{kms06} J. Kubo, E. Ma, and D. Suematsu, Phys. Lett. {\bf B642}, 18 
(2006) [hep-ph/0604114].
\bibitem{b18} M. Becker, arXiv:1806.08579 [hep-ph].
\bibitem{ck18} M. Chianese and S. F. King, JCAP {\bf 1809}, 027 (2018) 
[arXiv:1806.10606].
\bibitem{fm18} P. Fileviez Perez and C. Murgui, Phys. Rev. {\bf D98}, 
055008 (2018) [arXiv:1803.07462].
\bibitem{betal18} S. Bhattacharya {\it et al.}, JHEP {\bf 1812}, 007 (2018) 
[arXiv:1806.00490].
\bibitem{hjmw10} L. J. Hall, K. Jedamzik, J. March-Russell, and S. M. West, 
JHEP {\bf 1003}, 080 (2010) [arXiv:0911.1120].
\bibitem{ac13} G. Arcadi and L. Covi, JCAP {\bf 1308}, 005 (2013) 
[arXiv:1305.6587].

\end{thebibliography}

\end{document}